# Progress and opportunities in modelling environmentally assisted cracking

Emilio Martínez-Pañeda

Department of Civil and Environmental Engineering, Imperial College London, London SW7 2AZ, UK

**Abstract**

Environmentally assisted cracking phenomena are widespread across the transport, defence, energy and construction sectors. However, predicting environmentally assisted fractures is a highly cross-disciplinary endeavour that requires resolving the multiple material-environment interactions taking place. In this manuscript, an overview is given of recent breakthroughs in the modelling of environmentally assisted cracking. The focus is on the opportunities created by two recent developments: phase field and multi-physics modelling. The possibilities enabled by the confluence of phase field methods and electro-chemo-mechanics modelling are discussed in the context of three environmental assisted cracking phenomena of particular engineering interest: hydrogen embrittlement, localised corrosion and corrosion fatigue. Mechanical processes such as deformation and fracture can be coupled with chemical phenomena like local reactions, ionic transport and hydrogen uptake and diffusion. Moreover, these can be combined with the prediction of an evolving interface, such as a growing pit or a crack, as dictated by a phase field variable that evolves based on thermodynamics and local kinetics. Suitable for both microstructural and continuum length scales, this new generation of simulation-based, multi-physics phase field models can open new modelling horizons and enable *Virtual Testing* in harmful environments.

## 1. Introduction

Engineering failures are most often the result of the combined actions of mechanical loading and environmental material degradation. While a good understanding of how materials and structures behave when subjected to an applied load has been gained through the years, the prediction of material-environment interactions continues to be an elusive goal. The integrity and durability of engineering infrastructure are compromised by multi-physics phenomena such as stress corrosion cracking, hydrogen embrittlement, and corrosion fatigue. These environmentally assisted cracking phenomena are the primary cause of failure of metallic and metallic-reinforced structures in a wide range of industries, from power generation to construction [1]. Predicting environmentally assisted fractures requires highly inter-disciplinary endeavours, crossing the boundaries of materials science, electrochemistry and mechanics. Moreover, the understanding of these environment-material interactions is significantly hindered by their underlying complexity and the scientific challenges associated with them. The nature of environmentally assisted fractures can vary significantly from one material to another, or even between different environments within the same material, hindering a critical evaluation of the different mechanisms at play. As a consequence, the majority of the work conducted in the area of environmentally assisted cracking has been experimental, with modelling efforts traditionally restricted to the development of analytical, phenomenological models [2]. While phenomenological models can be useful in providing qualitative insight, they have limited applicability, as they require extensive calibration and have a regime of relevance limited to scenarios resembling the calibration schemes. Given the challenges associated with conducting experiments in conditions equivalent to those of engineering practice, there is a strong need to develop models that can explicitly resolve the physics, minimising the number of assumptions. For example, models that do not resolve the evolving topology of defects require strong assumptions to quantify their role on the local chemistry and the electrolyte flow. Recent progress in computer-based capabilities has opened the door to the development of simulation- and physics-based predictive models that can prevent catastrophic failures and optimise maintenance and design.

Two recent developments have arguably brought a step-change to the modelling of environmentally assisted fracture. First, larger computer resources and new algorithms enable simulating coupled physical processes such as the transport of various dilute species (chemistry), the distribution of the electric field and current density (electrochemistry) and the deformation of materials (mechanics), so-called multi-physics modelling [3,4]. It is now possible to simulate the concurrent physical processes occurring not only within the material but also in the environment. For example, one can model chemical reactions and ion transport in an aqueous electrolyte, while solving for lattice diffusion, deformation and fracture in the solid metal. Here, multi-physics refers to inter-physics coupling; i.e., not only the concurrent modelling of physical phenomena but also the development of universal relationships between the phenomena at play. The second element paving the way for a new generation of simulation-based models is the development of phase field algorithms, which enable researchers to computationally track evolving interfaces [5,6]. Localised corrosion and fracture are interfacial problems in which the challenge lies in predicting how the boundary between the solid metal and the electrolyte (or the cracked region) changes in time. Modelling morphological changes in interfaces is a well-known computational and mathematical challenge. It requires defining boundary conditions on a moving boundary and manually adjusting the interface topology with arbitrary criteria when merging or division occurs. Phase field models overcome these challenges by smearing the interface over a diffuse region, using an auxiliary phase field variable $\phi$. As shown in Fig. 1, this phase field variable takes two distinct magnitudes in each of the phases (e.g., 0 and 1) and exhibits a smooth change between these values near the interface. The temporal evolution of the phase field $\phi$ is described by a differential equation that can be readily solved by using established numerical methods, enabling the simulation of complex interface evolution phenomena [7–9]. Despite being relatively new, the phase field modelling paradigm has become the method of choice for simulating microstructural evolution in the condensed matter community. The change in shape and size of microstructural features such as grains can be predicted by defining the evolution of the phase field in terms of other fields (temperature, concentration, strain, etc.) through a thermodynamic free energy [10]. More relevant, the phase field paradigm has been recently extended to model fracture [11–13] and corrosion [14,15].

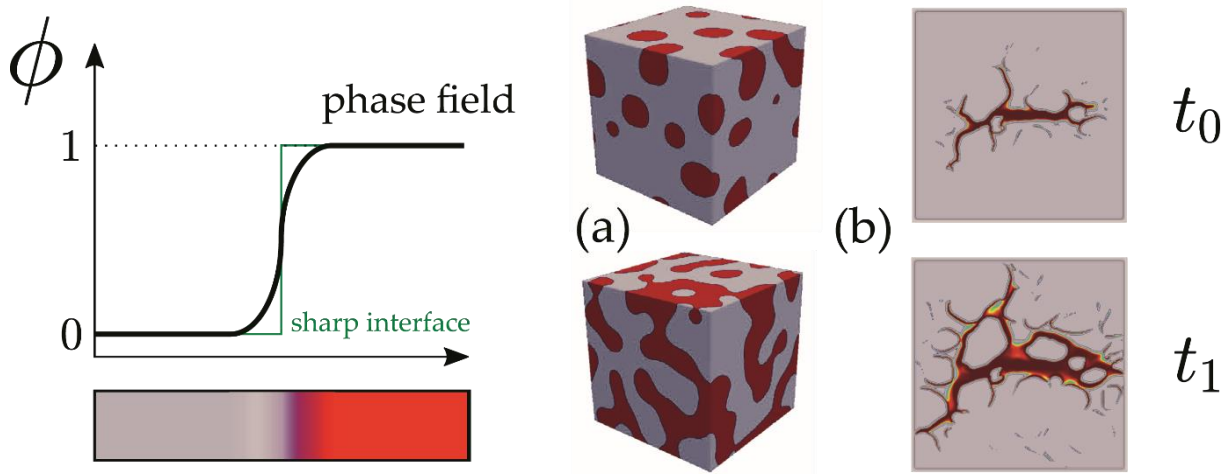

Fig. 1. Tracking interfaces implicitly using an auxiliary phase field $\phi$. Examples capturing (a) microstructural evolution [16], and (b) the propagation of cracks [17], for two instants in time ($t_0$, $t_1$). Adapted from [13].

The goal of this manuscript is to highlight recent breakthroughs in computer modelling of environmentally assisted cracking and to outline future opportunities, as enabled by multi-physics modelling, phase field and other recent methodological advances. The paper will separately address the progress and opportunities in three particularly relevant environmentally assisted cracking phenomena: hydrogen embrittlement (Section 2), pitting and stress corrosion cracking (Section 3), and corrosion fatigue (Section 4). Finally, concluding remarks are given in Section 5.

## 2. Hydrogen embrittlement

When exposed to hydrogen, metals experience a dramatic reduction in strength, ductility, fracture toughness and fatigue resistance, a phenomenon referred to as *hydrogen embrittlement* [18,19]. The problem is becoming increasingly important due to the higher susceptibility of modern, high-strength alloys; hydrogen can reduce the toughness of high-strength steels by 90% [20-23], compromising decades of metallurgical progress. Hydrogen embrittlement is pervasive in numerous applications across engineering sectors, such as bridges [24], buildings [25], cars [26], trains [27], wind turbines [28] and aeroplanes [29]. Moreover, hydrogen embrittlement could jeopardise the future that hydrogen holds as the energy carrier of the future. RILEM's

CCH Technical Committee[1] investigates hydrogen assisted failures in prestressed and galvanised tendons, radioactive waste packages, ground anchors and bridge straps, among others.

Hydrogen ingress into a metal can happen during its initial forming, during the coating or plating of a protective layer, through exposure to hydrogen or hydrogen-containing molecules in the air, soil or water, or through corrosion processes. Hydrogen diffuses through metal crystal lattices and accumulates in areas of high hydrostatic stress, where the associated expansion of the lattice results in a higher local hydrogen solubility [30]. Hydrogen atoms sit either at interstitial lattice sites or at pre-existing defects like grain boundaries or dislocations, which act as microstructural *traps* [31–33]. Hydrogen assisted fracture takes place when a critical combination of hydrogen concentration and applied mechanical load is reached, through mechanisms that vary among material systems and environments [34]–[38]. Recent years have seen a surge in the development of computer-based multi-physics (chemo-mechanics) models for predicting hydrogen assisted cracking (see, e.g., [39–43] and Refs. therein). As described below, recent work in two lines of research has opened new promising horizons for the modelling and prediction of hydrogen embrittlement.

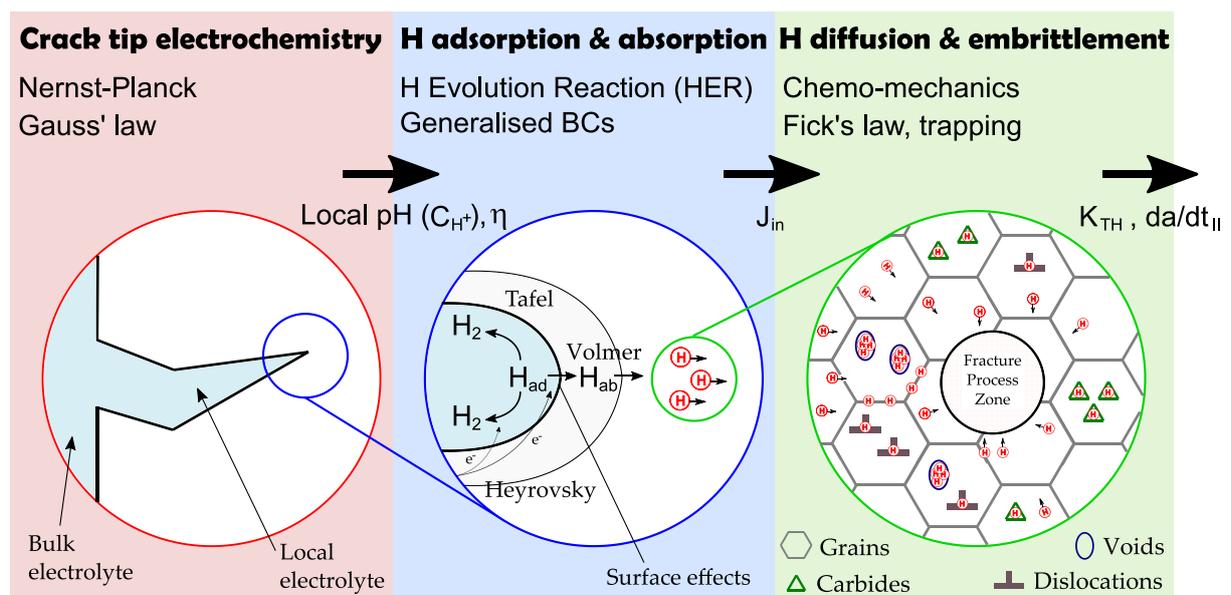

Fig. 2. Stages involved in the hydrogen uptake and embrittlement processes: (1) electrochemistry (red), (2) electrochemistry-diffusion interface (blue), and (3) diffusion and trapping (green).

The first line of research is related to quantifying hydrogen ingress in metals, a relatively straightforward task in the presence of pure $H_2$ gas (Sievert's law), but a complex endeavour in other environments (e.g., water vapour or aqueous electrolytes). The three main stages involved in hydrogen uptake are shown in Fig. 2. The vast majority of existing models resolve stage 3; the coupled deformation-diffusion behaviour of the metal-hydrogen system is simulated and connected to a fracture criterion that is a function of mechanical fields and hydrogen concentration. The predictions of these chemo-mechanical models are very sensitive to a key input: the content of dissolved atoms at the crack surfaces. Typically, a fixed hydrogen concentration is prescribed at the boundary, but its quantification remains a challenge [44-45].

---
[1] RILEM CCH Technical Committee: Stress Corrosion Cracking and Hydrogen Embrittlement of Concrete-Reinforcing Steels.

Recent efforts have been devoted to tackling stage 2: resolving the electrochemical-diffusion interface. Models have been presented that provide as input the lattice chemical potential [46–49], a more rigorous approach than prescribing a constant hydrogen concentration. More importantly, a new generalised formulation has been proposed to resolve the electrochemistry-diffusion interface, enabling quantifying the flux of hydrogen from the local pH and overpotential [50]. These generalised boundary conditions not only provide a more accurate description of hydrogen ingress but also open the door to the development of a new generation of fully predictive electro-chemo-mechanical models, which explicitly resolve the crack tip electrochemistry (stages 1 to 3). This is important because the composition of the electrolyte within pits, cracks and other occluded environments is considerably different from that of the bulk solution [51]. For example, the pH can drop from 9 in the bulk to 2 inside of a defect, due to the combination of the separation of the locations of the anodic and cathodic reactions and the hydrolysis of the metal cations produced by dissolution; this results into an increase in the hydrogen ion concentration. Thus, there is a need to explicitly simulate the electrolyte behaviour (Nernst-Planck, Gauss' law) and to faithfully capture the morphology of evolving defects at the electrolyte-metal interface.

The second element that has opened new modelling horizons is the extension of the phase field paradigm to model hydrogen assisted fracture and fatigue. Phase field fracture models are enjoying significant popularity due to their ability to (i) capture complex cracking phenomena such as crack merging and branching, in arbitrary geometries and dimensions [52]–[54]; (ii) provide a platform for the simple yet rigorous fracture thermodynamics principles first presented by Griffith [55]; and (iii) simulate coupled multi-physics fracture problems efficiently and without convergence problems. The phase field $\phi$ varies from 0 to 1, akin to a damage variable, and mesh objectivity is guaranteed by the non-locality of the model, with a length scale $\ell$ parameter defined to govern the smearing of the crack. A phase field fracture formulation for hydrogen embrittlement was presented by Martínez-Pañeda *et al.* [56] in 2018, and the field has then gained significant traction (see, e.g. [57]–[60] and Refs. therein). The phase field evolution law is the result of considering the variational form of Griffith's energy balance – a crack will grow when the energy stored in the solid is sufficient to overcome the energy required to create two new surfaces; a competition between the strain energy density $\psi(\boldsymbol{\varepsilon})$, which is a function of the strain tensor $\boldsymbol{\varepsilon}$, and the material toughness $G_c$. The role of hydrogen in facilitating fracture is captured by introducing a hydrogen degradation function $f(C)$, which reduces the toughness of the material, as observed experimentally. The phase field evolution law then reads:

$$G_c f(C) \left(\frac{\phi}{\ell} - \ell \nabla^2 \phi\right) - 2(1-\phi)\psi(\boldsymbol{\varepsilon}) = 0 \qquad (1)$$

The model is universal and $f(C)$ can be defined in a phenomenological manner (calibrating with experiments) or chosen to accommodate any mechanistic interpretation. For example, the degradation function has been defined based on atomistic calculations of surface energy susceptibility to hydrogen coverage, upon the assumption of embrittlement due to atomic decohesion [56]. Among others, phase field formulations for hydrogen assisted fracture have been used to (i) assess the suitability of Slow Strain Rate Tests (SSRTs), the testing configuration most widely used to evaluate hydrogen susceptibility [61], (ii) predict the hydrogen-enhancement of fatigue crack growth rates, providing *Virtual S-N curves* [62]; and (iii) conduct virtual laboratory and field experiments [63,64]. Representative results are shown in Fig. 3. Failure of structural elements such as screw anchors and pipelines can now be predicted, as a function of the material properties, the environment and the loading conditions. The ability of phase field methods to simulate the complex conditions inherent to engineering practice has opened the door to the use of simulation-based paradigms, such as *Virtual Testing* or *Digital Twins*, in hydrogen-sensitive applications. Finally, it is important to emphasise that the phase field method can be used to predict the evolution of defects of any size and geometry, not only sharp cracks. This not only provides a more realistic assessment but also enables to completely resolve the multi-physical nature of the problem, *via* the coupling with the electro-chemo-mechanics models described above.

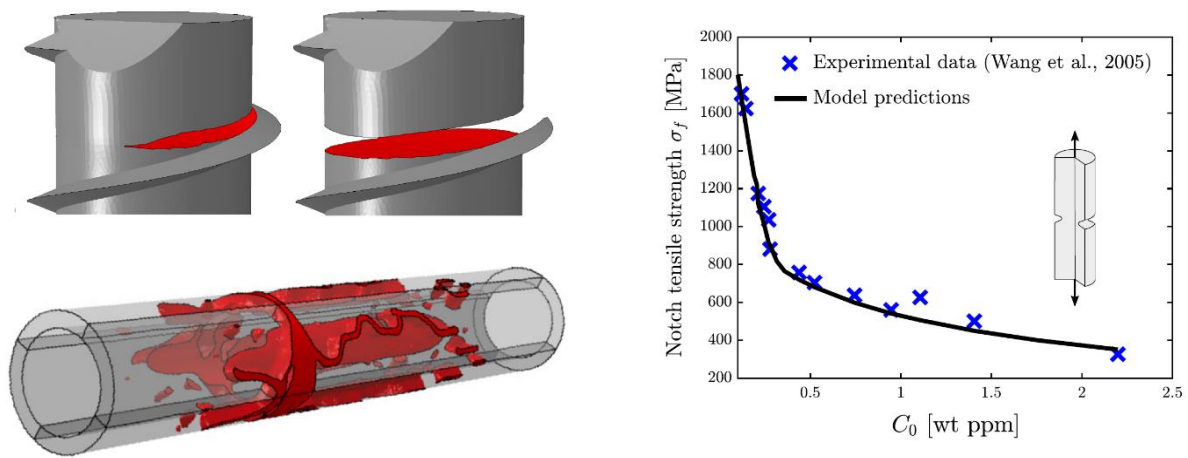

Fig. 3. Phase field modelling of hydrogen embrittlement in a screw anchor, a pipeline with internal defects and in cylindrical notched bars. The results reported belong to Refs. [63,64].

## 3. Pitting and stress corrosion cracking

Environmentally assisted fractures can be cathodic-driven and associated with the ingress and transport of hydrogen, as covered in Section 2, or driven by material dissolution and the anodic corrosion reaction at the crack tip, as addressed in this Section. While the focus on this Section is on anodic dissolution-controlled Stress Corrosion Cracking (SCC), the conclusions and formulations can be readily generalised to other SCC phenomena, such as hydrogen embrittlement and sulfide stress cracking (SSC). Environmentally assisted fracture mechanisms are not mutually exclusive and, for example, hydrogen assisted fractures are largely facilitated by the presence of pits and other stress concentrators, which have been triggered due to material dissolution. A large body of literature has emerged to phenomenologically capture anodic-driven SCC, relevant works include the active path dissolution model [65], the surface mobility model [66], the coupled environment fracture model [66,67], and models grounded on the film rupture–dissolution–repassivation mechanism [68,69]. Common to many of these models is the assumption that a protective film forms under passivation conditions and that film-rupture is a necessary condition for damage to occur. The damage process itself can be limited to metallic dissolution or extended to accommodate other mechanisms, as it is often assumed in the case of transgranular SCC, to rationalise the differences between crack growth and dissolution rates [70,71]. There is a need to shed light on these hypotheses by simulating explicitly the underlying physics and extending modelling approaches to capture the important role of the defect geometry in driving local chemistry and dissolution rates. The phase field paradigm provides a suitable framework to achieve this.

Phase field modelling has very recently been extended to corrosion [14-15]. The phase field $\phi$ is used to describe the evolution of the solid metal-aqueous electrolyte interface, enabling to capture pitting, the pit-to-crack transition and stress corrosion cracking – see Fig. 4. In its simplest form, the free energy of the system $F$ is decomposed into the bulk $F_b$ and interface $F_i$ contributions as follows,

$$F = F_b + F_i = \int \left(f(c,\phi) + f_i(\nabla\phi)\right) dV = \int \left(f(c,\phi) + \frac{\alpha}{2}(\nabla\phi)^2\right) dV \quad (2)$$

where $c$ is the normalised concentration of dissolved metal ions, $f$ is the local bulk free energy density and $f_i$ is the interface energy density, which is a function of the phase field gradient and the gradient energy coefficient $\alpha$. The governing equations for the phase field and the concentration of dissolved ions are then derived by minimising the free energy of the system $F$:

$$\frac{\partial \phi}{\partial t}(\boldsymbol{x},t) = -L\frac{\delta F}{\delta \phi} = -L\left(\frac{\partial f}{\partial \phi} - \alpha\nabla^2\phi\right) \quad (3)$$

$$\frac{\partial c}{\partial t}(\boldsymbol{x},t) = \nabla \cdot M\nabla \frac{\delta F}{\delta c} = \nabla \cdot M\left(\nabla \frac{\partial f}{\partial c}\right) \quad (4)$$

Here, $L$ is the interface kinetics parameter and $M$ is the diffusion mobility coefficient, which can be re-written as $M = D/(2A)$, where $D$ is the diffusion coefficient and $A$ is a temperature-dependent free energy density proportionality constant.

The bulk free energy can be defined considering that the concentration at any point is evaluated as the weighted sum of the solid and liquid concentrations. Such, that for a double well potential $g(\phi) = \phi^2(1-\phi)^2$ with weight $w$, it reads:

$$f(c,\phi) = h(\phi)f_S(c_S) + \left(1 - h(\phi)\right)f_L(c_L) + wg(\phi) \quad (6)$$

where $h(\phi)$ is an interpolation function that must satisfy $h(0) = 0$ and $h(1) = 1$.

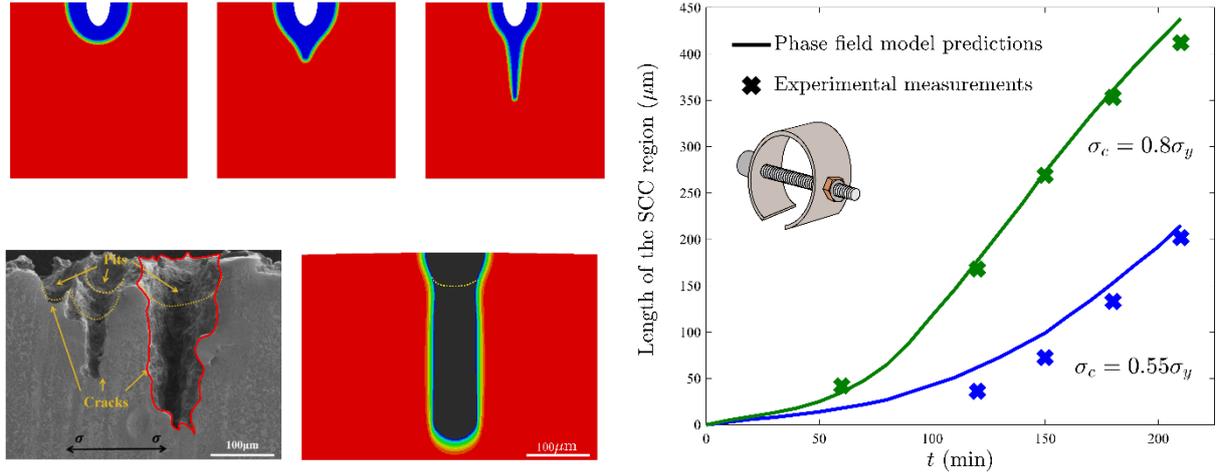

Fig. 4. Phase field modelling of the processes of pitting, pit-to-crack transition and SCC, including experimental validation. The results reported belong to Ref. [15].

It remains to define the constitutive choices of the bulk free energy density for the liquid $f_L$ and solid $f_S$ phases. Based on the Kim–Kim–Suzuki (KKS) model [73], the material is regarded as a mixture of two coexisting phases at each point, and a local equilibrium between the two phases is always assumed:

$$c = h(\phi)c_S + \big(1 - h(\phi)\big)c_L \quad \text{and} \quad \frac{\partial f_S(c_S)}{\partial c_S} = \frac{\partial f_L(c_L)}{\partial c_L} \quad (5)$$

where $c_S$ and $c_L$ respectively denote the normalised concentrations in the solid and the liquid. Assuming a similar $A$ for the solid and liquid phases, the free energy densities read:

$$f_S = A(c_S - c_{Se})^2 = A(c_S - 1)^2 \quad \text{and} \quad f_L = A(c_L - c_{Le})^2 = A\left(c_L - \frac{c_{sat}}{c_{solid}}\right)^2 \quad (6)$$

where $c_{Se}$ and $c_{Le}$ are the normalised equilibrium equations in the solid and liquid phases, $c_{solid}$ is the concentration of atoms in the metal and $c_{sat}$ is the saturation concentration, at which corrosion becomes diffusion-controlled. Inserting these constitutive choices into (3)-(4) delivers a model capable of predicting the evolution of the corrosion front, for arbitrary geometries and dimensions, based on thermodynamic and kinetic principles. Importantly, the model has been recently extended to incorporate film rupture and repassivation, as well as the influence of mechanical fields in corrosion kinetics [15]. Another important strength of phase field corrosion is the possibility of easily coupling the method with existing models for electrochemistry (ionic transport, current density) and microstructure (e.g., to capture the role of crystallographic orientations) [73-74]. Thus, phase field opens new horizons in the modelling of corrosion and can provide key insight: from shedding light into the role of dynamic plastic straining in pit growth [76] to predicting corrosion induced cracking in reinforced concrete [77]. The confluence of phase field corrosion and recent progress in high-resolution measuring techniques, such as X-ray tomography [77-78], sets the basis for the development of a new generation of validated, simulation-based models capable of predicting local corrosion attack and its consequences.

## 4. Corrosion fatigue

Corrosion fatigue is a well-known structural integrity concern [80]. In many applications, metals are exposed to harmful environments and alternating mechanical loads during their service life. For example, corrosion fatigue is the limit state consideration for the design and in-service life of offshore wind structures. However, corrosion fatigue is also known to be a particularly complex phenomenon – it involves the interaction of two damage mechanisms (corrosion and fatigue) that are very complex on their own and, individually, not yet fully understood. The evolution of corrosion fatigue damage is typically divided into four regimes [81]: (1) surface film rupture, (2) pit growth, (3) pit-to-crack transition, and (4) fatigue crack growth. The four stages can be captured with phase field, either by using a generalised phase field that accounts for both pitting and fracture [82] or by taking a multi-phase field approach [83]. As sketched in Fig. 5 and discussed above, phase field predictions of pitting and cracking can also be coupled to electrolyte modelling, to capture the important role that the defect shape plays in the local chemistry. Moreover, phase field fracture models have also been extended to fatigue [84], enabling to naturally *predict* S-N curves, the Palmgren–Miner rule and Paris' law behaviour. Very recently, phase field fatigue has been extended to account for environmental effects and deliver *Virtual* S-N curves that can replace empirical knock-down factors [62].

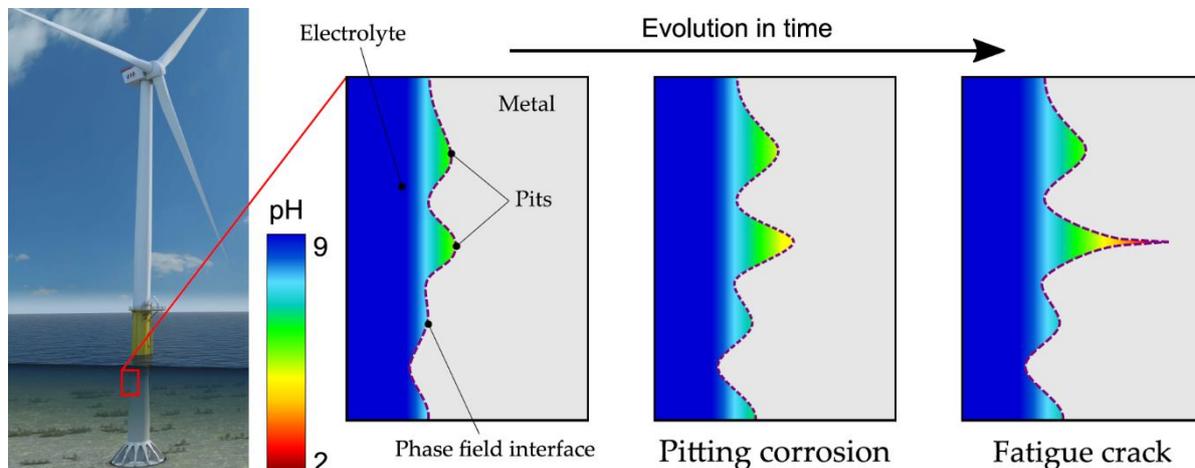

Fig. 5. Capturing morphological changes in the electrolyte-metal interface using phase field: from pitting corrosion to the initiation and growth of fatigue cracks.

Phase field models open a path for predicting corrosion fatigue, a long-standing challenge. However, this path is not free of hurdles and, while many effects are now simulatable, the computational cost of resolving all the physical mechanisms at play must be balanced with their relevance in the damage process. Delivering reliable predictions at engineering scales requires determining what are first order and second order effects, and phase field modelling can provide a virtual platform to assess the weight of the various mechanisms taking place. For example, most corrosion models neglect the contribution of convection to mass transport upon the assumption of a low flow rate; however, electrolyte flow can be significant in fatigue [85]. Phase field modelling can also be used to quantify the role of other important effects, such as the electrochemical crack size effect [86]. A myriad of opportunities can be exploited by combining phase field and multi-physics modelling, and increasing computer power will pave the way for a new era of simulation-based structural integrity assessment of structures and materials in aggressive environments.

## 5. Concluding remarks

Two recent developments, phase field and multi-physics modelling, have opened new horizons in the modelling of material-environment interactions and their implications on structural integrity. Specifically, this perspective article focuses on recent progress and opportunities in relation to three phenomena of particular engineering interest: hydrogen embrittlement, localised corrosion and corrosion fatigue. Phase field modelling enables capturing, in both 2D and 3D, the evolving morphology of defects of arbitrary geometry. The propagation of cracks can be predicted based on the thermodynamics of fracture and the growth of pits is simulated based on thermodynamic and kinetic laws. The coupling of phase field modelling with multi-physics simulations enables capturing, as a function of time, the interplay between defect morphology, local chemistry, dissolution kinetics and cracking. Local chemical reactions, ionic transport, hydrogen diffusion and mechanical straining can be simulated in the vicinity of an interface that evolves as dictated by corrosion and fracture processes. The continuum-like nature of these models enables dealing with the large space and time scales relevant to engineering practice; there is an opportunity to extend the success of *Virtual Testing* in the automotive and aerospace sectors to applications involving material-environmental interactions, from offshore wind turbine monopiles to reinforced concrete structures.

## Acknowledgements

I am very grateful and honoured to receive the Gustavo Colonnetti Medal from RILEM. My work has always been a team effort and thus I would like to thank my collaborators, students, and postdocs for making this possible. I would also like to acknowledge financial support from the Engineering and Physical Sciences Research Council (EPSRC) through grant EP/V009680/1 and from the Royal Commission for the 1851 Exhibition (RF496/2018).